\documentclass[aps,prl,showpacs,twocolumn,floats,epsfig]{revtex4}
\usepackage{amssymb}
\usepackage{amsbsy}
\usepackage{amsmath}
\usepackage{epsfig}

\begin{document}

\title {Tunneling conductance of graphene NIS junctions }

\author{Subhro Bhattacharjee$^{(1)}$ and K. Sengupta$^{(2)}$}

\affiliation{$^{(1)}$CCMT, Department of Physics, Indian Institute of Science, Bangalore-560012, India.\\
$^{(2)}$TCMP division, Saha Institute of Nuclear Physics, 1/AF
Bidhannagar, Kolkata-700064, India. }

\date{\today}

\begin{abstract}

We show that in contrast to conventional normal
metal-insulator-superconductor (NIS) junctions, the tunneling
conductance of a NIS junction in graphene is an oscillatory function
of the effective barrier strength of the insulating region, in the limit
of a thin barrier. The amplitude of these oscillations are maximum for
aligned Fermi surfaces
of the normal and superconducting regions and vanishes for large Fermi surface
mismatch. The zero-bias tunneling conductance, in sharp contrast to
its counterpart in conventional NIS junctions, becomes maximum for a
finite barrier strength. We also suggest experiments to test these
predictions.

\end{abstract}


\maketitle

Graphene, a two-dimensional single layer of graphite, has been
recently fabricated by Novoselov {\it et.\,al.} \cite{nov1}. This
has provided an unique opportunity for experimental observation of
electronic properties of graphene which has attracted theoretical
attention for several decades \cite{oldref}. In graphene, the energy
bands touch the Fermi energy at six discrete points at the edges of
the hexagonal Brillouin zone. Out of these six Fermi points, only
two are inequivalent; they are commonly referred to as $K$ and $K'$
points \cite{ando1}. The quasiparticle excitations about these $K$
and $K'$ points obey linear Dirac-like energy dispersion. The
presence of such Dirac-like quasiparticles is expected to lead to a
number of unusual electronic properties in graphene including
relativistic quantum hall effect with unusual structure of Hall
plateaus \cite{shar1}. Recently, experimental observation of the
unusual plateau structure of the Hall conductivity has confirmed
this theoretical prediction \cite{nov2}. Further, as suggested in
Ref.\ \cite{geim1},the presence of such quasiparticles in graphene
provides us with an experimental test bed for Klein paradox \cite{klein1}

Another, less obvious but nevertheless interesting, consequence of
the existence Dirac-like quasiparticles can be understood by
studying tunneling conductance of a normal metal-superconductor (NS)
interface of graphene \cite{beenakker1}. Graphene is not a
natural superconductor. However, superconductivity can be induced in a
graphene layer in the presence of a superconducting electrode near it
via proximity effect \cite{volkov1,beenakker1,beenakker2}. It has been
recently predicted \cite{beenakker1} that a graphene NS junction, due to the Dirac-like
energy spectrum of its quasiparticles, can exhibit specular Andreev
reflection in contrast to the usual retro reflection observed in
conventional NS junctions \cite{andreev1,tinkham1}. Such specular
Andreev reflection process leads to qualitatively different
tunneling conductance curves compared to conventional NS junctions
\cite{beenakker1}. However, the effect of the
presence of a potential barrier between the normal and
superconducting regions in graphene NS junction has not been studied
so far.

In this letter, we study the tunneling conductance of a normal
metal-insulator-superconductor (NIS) junction of graphene in the
limit of thin barrier. We show that in contrast to the
conventional NIS junctions, the tunneling conductance of a graphene
NIS junction is an oscillatory function of the effective barrier
strength. The amplitude of these
oscillations is maximum for aligned Fermi surfaces of the normal and
superconducting regions and vanishes for large Fermi surface
mismatch. In particular, we point out that the zero-bias
conductance, in complete contrast to its behavior in conventional
NIS junctions, reaches its maximum value for a finite
barrier strength. By using the fact that the effective barrier
height can be tuned experimentally by
changing a gate voltage \cite{nov2,geim1}, we suggest an
experimental setup where these effects can be observed. We also
point out that our analysis reproduces the results of previous work
on graphene NS junctions as a special case \cite{beenakker1}.

We consider a NIS junction in a graphene sheet occupying the $xy$
plane with the normal region extending $x=-\infty$ to $x=-d$ for all
$y$. The region I, modeled by a barrier potential $V_0$,
extends from $x=-d$ to $x=0$ while the
superconducting region occupies $x\ge 0$. Such a local
barrier can be implemented by either using the electric field effect
or local chemical doping \cite{geim1,nov2}. The region $x \ge 0$ is
to be kept close to an superconducting electrode so that
superconductivity is induced in this region via proximity effect
\cite{volkov1,beenakker1}. In the rest of this work, we shall assume
that the barrier region has sharp edges on both sides. This condition
requires that $d \ll 2\pi/k_F$, where $k_F$ is the Fermi
wave-vector for graphene, and can be realistically created in
experiments \cite{geim1}. The NIS junction can then be described by
the Dirac-Bogoliubov-de Gennes (DBdG) equations \cite{beenakker1}
\begin{eqnarray}
&&\left(\begin{array}{cc}
    {\mathcal H}_{a}-E_F + U({\bf r}) & \Delta ({\bf r}) \\
     \Delta^{\ast}({\bf r}) & E_F - U({\bf r})-{\mathcal H}_{a}
    \end{array}\right) \psi_{a}   = E \psi_{a}. \nonumber\\
\label{bdg1}
\end{eqnarray}
Here, $\psi_a = \left(\psi_{A\,a}, \psi_{B\,a}, \psi_{A\,{\bar
a}}^{\ast}, -\psi_{B\,{\bar a}}^{\ast}\right)$ are the $4$ component
wavefunctions for the electron and hole spinors, the index $a$
denote $K$ or $K'$ for electron/holes near $K$ and $K'$ points,
${\bar a}$ takes values $K'(K)$ for $a=K(K')$, $E_F$ denote the
Fermi energy, $A$ and $B$ denote the two inequivalent sites in the
hexagonal lattice of graphene, and the Hamiltonian ${\mathcal H}_a$
is given by
\begin{eqnarray}
{\mathcal H}_a &=& -i \hbar v_F \left(\sigma_x \partial_x + {\rm
sgn}(a) \sigma_y
\partial_y \right). \label{bdg2}
\end{eqnarray}
In Eq.\ \ref{bdg2}, $v_F$ denotes the Fermi velocity of the
quasiparticles in graphene and ${\rm sgn}(a)$ takes values $\pm$ for
$a=K(K')$.

The pair-potential $\Delta({\bf r})$ in Eq.\ \ref{bdg1} connects the
electron and the hole spinors of opposite Dirac points. We have
modeled the pair-potential as $\Delta({\bf r}) = \Delta_0
\exp(i\phi) \theta(x)$, where $\Delta_0$ and $\phi$ are the
amplitude and the phase of the induced superconducting order
parameter respectively and $\theta$ is the Heaviside step function.
The potential $U({\bf r})$ gives the relative shift of Fermi
energies in normal, insulating and superconducting regions of
graphene and can be modeled as $U({\bf r}) = -U_0 \theta(x) + V_0
\theta(-x) \theta(x+d)$. At this stage, we introduce the
dimensionless barrier strength
\begin{eqnarray}
\chi = V_0 d/\hbar v_F, \label{barstr}
\end{eqnarray}
which is going to play a key role in the subsequent analysis. In
particular, we define a thin barrier as one with $V_0 \rightarrow
\infty$ and $d\rightarrow0$ such that $\chi$ remains finite. For the
NS junction studied in Ref.\ \cite{beenakker1},
$\chi=0$. The gate potential $U_0$ can be used to tune the Fermi
surface mismatch between the normal and the superconducting regions.
Notice that the mean-field conditions for superconductivity is
satisfied as long as $\Delta_0 \ll (U_0 +E_f)$; thus, in principle,
for large $U_0$ one can have regimes where $\Delta_0 \ge E_f$
\cite{beenakker1}.

Eq.\ \ref{bdg1} can be solved in a straightforward manner to yield
the wavefunction $\psi$ in the normal, insulating and the
superconducting regions. In the normal region, for electron and
holes traveling the $\pm x$ direction with a transverse momentum
$k_y=q$ and energy $\epsilon$, the wavefunctions are given by
\begin{eqnarray}
\psi_N^{e \pm} &=&  \left(1,\pm e^{\pm i \alpha},0,0\right) \exp
\left[i \left(\pm k_{n} x + q y\right) \right], \nonumber\\
\psi_N^{h \pm} &=& \left(0,0,1,\mp e^{\pm i \alpha'}\right) \exp
\left[i \left(\pm k'_{n} x + q y \right)\right],
\nonumber\\
\sin(\alpha) &=& \frac{\hbar v_F q}{\epsilon +E_F},  \quad
\sin(\alpha') = \frac{\hbar v_F q}{\epsilon - E_F}, \label{wavenorm}
\end{eqnarray}
where for the electron wavefunctions $k_n = \left(\epsilon + E_F
\right) \cos(\alpha)/\hbar v_F$ and $\alpha$ is the angle of
incidence of the electron. Similarly for the hole wavefunctions,
$k'_n = \left(\epsilon - E_F \right) \cos(\alpha')/\hbar v_F$ with
angle of incidence $\alpha'$. Note that for an Andreev process to
take place, the maximum angle of incidence for an electron is given
by $\alpha_{c} = \arcsin\left[\left|\epsilon -
E_F\right|/\left(\epsilon + E_F\right)\right]$ \cite{beenakker1}.

In the barrier region, one can similarly obtain $\psi_B^{e \pm} =
\left(1,\pm e^{\pm i \theta},0,0\right) \exp \left[i\left(\pm k_{b}
x + q y \right)\right]$ and  $\psi_B^{h \pm} = \left(0,0,1,\mp
e^{\pm i \theta'}\right) \exp \left[i \left(\pm k'_{b} x + q y
\right)\right]$ for electron and holes moving along $\pm x$. Here
the angle of incidence of the electron(hole) $\theta(\theta')$ is
given by $\sin\left[\theta(\theta')\right] = \hbar v_F
q/\left[\epsilon +(-)(E_F-V_0)\right]$ and $k_b (k'_b) =
\left[\epsilon -(+) (E_F-V_0) \right]
\cos\left[\theta(\theta')\right]/\hbar v_F$. Note that in the limit
of thin barrier, $\theta, \theta' \rightarrow 0$ and $k_b d, k'_b d
\rightarrow \chi$.

In the superconducting region, the BdG quasiparticles are mixtures
of electron and holes. Consequently, the wavefunctions of the BdG
quasiparticles moving along $\pm x$ with transverse momenta $q$ and
energy $\epsilon$ has the form
\begin{eqnarray}
\psi_S^{\pm} &=& \left( e^{\mp i \beta}, \mp e^{\pm i \left(\gamma
-\beta\right)}, e^{-i\phi},\mp e^{i\left(\pm \gamma -\phi \right)}
\right), \nonumber\\
&& \times \exp\left[ i\left(\pm k_s x +q y\right) - \kappa x\right],
\nonumber\\
\sin(\gamma) &=& \hbar v_F q/(E_F + U_0),
 \label{supwave}
\end{eqnarray}
where $k_s = \sqrt{\left[\left(U_0+E_F\right)/\hbar v_F\right]^2
-q^2}$ and $\gamma$ is the angle of incidence for the
quasiparticles. Here $\kappa^{-1} = (\hbar v_F)^2 k_s/\left[(U_0+E_F)
\Delta_0 \sin(\beta)\right]$ is the localization length and
$\beta$ is given by
\begin{eqnarray}
\beta &=& \cos^{-1} \left(\epsilon/\Delta_0\right) \quad {\rm if}
\left|\epsilon\right| < \Delta_0 ,\nonumber\\
&=& -i \cosh^{-1} \left(\epsilon/\Delta_0\right) \quad {\rm if}
\left|\epsilon\right| > \Delta_0, \label{betaeq}
\end{eqnarray}
Note that for $\left|\epsilon\right| > \Delta_0$, $\kappa$ becomes
imaginary and the quasiparticles can propagate in the bulk of the
superconductor.

Let us now consider an electron incident on the barrier from the
normal side with an energy $\epsilon$ and transverse momentum $q$.
The wave functions in the normal, insulating and superconducting
regions, taking into account both Andreev and normal reflection
processes, can then be written as \cite{tinkham1}
\begin{eqnarray}
\Psi_N &=& \psi_N^{e +}+r \psi_N^{e -} + r_A \psi_N^{h -}, \quad
\Psi_S = t \psi_S^{+}+ t' \psi_S^{-}, \nonumber\\
\Psi_B &=& p \psi_B^{e +}+q \psi_B^{e -} + m \psi_B^{h +} + n
\psi_N^{h -}, \label{wave2}
\end{eqnarray}
where $ r$ and $r_A$ are the amplitudes of normal and Andreev
reflections respectively, $t$ and $t'$ are the amplitudes of
electron-like and hole-like quasiparticles in the superconducting
region and $p$, $q$, $m$ and $n$ are the amplitudes of electron and
holes in the barrier. These wavefunctions must satisfy the
appropriate boundary conditions:
\begin{eqnarray}
\Psi_N |_{x=-d} &=& \Psi_B |_{x=-d},  \quad  \Psi_B |_{x=0} = \Psi_S
|_{x=0}. \label{bc1}
\end{eqnarray}
Notice that these boundary conditions, in contrast their
counterparts in standard NIS interfaces, do not impose any
constraint on derivative of the wavefunctions at the boundary. Thus
the standard delta function potential approximation for thin barrier
\cite{tinkham1} can not be taken the outset, but has to be taken at
the end of the calculation.

\begin{figure}
\rotatebox{0}{
\includegraphics*[width=7.0cm]{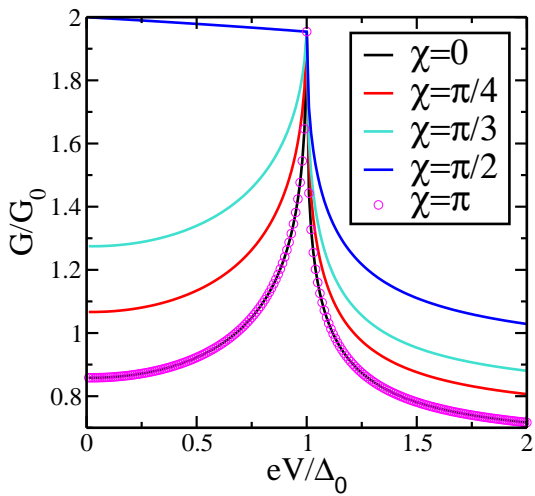}}
\caption{Plot of tunneling conductance of a NIS junction graphene as
a function of bias voltage for different effective barrier strengths
for $U_0=0$ and $\Delta_0=0.01 E_F$. Note that the curves for
$\chi=0$ (black line) and $\chi=\pi$(pink circles) coincide
reflecting $\pi$ periodicity.  } \label{fig1}
\end{figure}

Using the boundary conditions (Eq.\ \ref{bc1}), one can now solve
for the coefficients $r$ and $r_A$ in Eq.\ \ref{wave2}. After some
straightforward but cumbersome algebra, we find that in the limit of
thin barriers, the expressions for $r$ $t$, $t'$, and $r_A$ depend
on the dimensionless coefficient $\chi$ as
\begin{eqnarray}
r&=& \frac{\cos(\chi) \left(e^{i \alpha}-\rho\right) - i
\sin(\chi)\left(1-\rho e^{i \alpha} \right)}{\cos(\chi) \left(e^{-i
\alpha}+\rho\right) + i \sin(\chi)\left(1+\rho e^{-i \alpha}
\right)},\nonumber\\
 t' &=& \frac{\cos(\chi) \left(1+r\right) - i \sin(\chi)\left(e^{i
\alpha}-r e^{-i \alpha}\right)}{\Gamma e^{-i\beta} + e^{i
\beta}}, \quad t= \Gamma t',\nonumber\\
r_A &=& \frac{t'\left(\Gamma+1\right) e^{-i \phi}}{\cos(\chi) - i
e^{-i \alpha'} \sin(\chi)}, \label{coeff1}
\end{eqnarray}
where the parameters $\Gamma$ and $\rho$ can be expressed in terms
of $\gamma$, $\beta$, $\alpha$, and $\alpha'$ (Eqs.\ \ref{wavenorm},
\ref{supwave}, and \ref{betaeq}) as
\begin{eqnarray}
\Gamma &=& \frac{e^{-i\gamma} -\eta}{e^{i\gamma} +\eta}, \quad \eta
 = \frac{e^{- i \alpha'} \cos(\chi)  - i \sin(\chi)}{
\cos(\chi) - i e^{-i \alpha'} \sin(\chi)}, \nonumber\\
\rho &=& \frac{e^{-i(\gamma - \beta)} - \Gamma e^{i(\gamma -
\beta)}}{\Gamma e^{-i\beta} + e^{i \beta}}. \label{coeff2}
\end{eqnarray}
The tunneling conductance of the NIS junction can now be expressed
in terms of $r$ and $r_A$ by \cite{tinkham1}
\begin{eqnarray}
G(eV)&=& G_0 \int_0^{\alpha_c} \left(1-\left|r\right|^2 +
\left|r_A\right|^2 \frac{\cos(\alpha')}{\cos(\alpha)} \right)
\cos(\alpha) \, d\alpha, \nonumber \\ \label{tc1}
\end{eqnarray}
where $G_0 = 4e^2 N(eV)/h$ is the ballistic conductance of metallic graphene, $eV$
denotes the bias voltage, and $N(eV)= (E_f
+\epsilon)w/(\pi \hbar v_F)$ denotes the number of available
channels for a graphene sample of width $w$. Note that for $eV \ll
E_F$, $G_0$ is a constant. Eq.\ \ref{tc1} can be evaluated
numerically to yield the tunneling conductance of the NIS junction
for arbitrary parameter values.

Eqs.\ \ref{coeff1} and \ref{coeff2} represent the key result of this
work. From these equations, we find that in contrast to
conventional NIS junctions \cite{tinkham1}, both $r$ and
$r_A$ are {\it oscillatory functions of the effective barrier
potential $\chi$ for any angle of incidence $\alpha < \alpha_c$}.
Consequently, we expect $G(eV)$ (Eq.\ \ref{tc1})
to demonstrate oscillatory behavior as a function of $\chi$ with a
period $\pi$. We also note that our work reproduces the results
of Ref.\ \onlinecite{beenakker1} as a special case when $\chi=n \pi$
for any integer $n$.

\begin{figure}
\rotatebox{0}{
\includegraphics*[width=7.0cm]{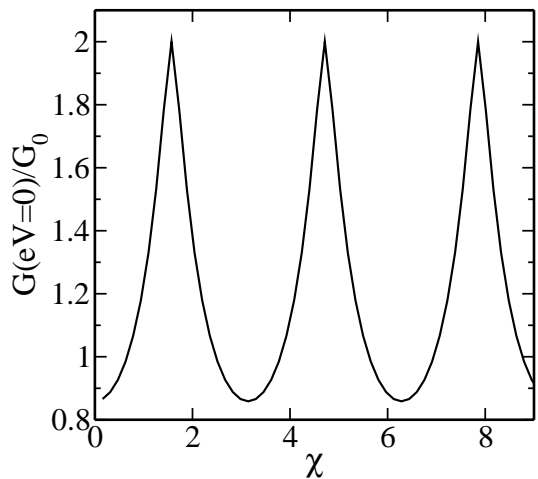}}
\caption{Plot of zero-bias tunneling conductance as a function of
effective barrier potential $\chi$ for $U_0=0$ and $\Delta_0=0.01
E_F$. The $\pi$ periodic oscillatory behavior with maxima at $\chi=(n+1/2)\pi$ is
to be contrasted with the monotonous decay of $G(eV=0)$
with increasing $\chi$ in conventional NIS junctions}
\label{fig2}
\end{figure}

Let us now consider the regime where the Fermi surfaces of the
normal metal and the superconductor is aligned ($U_0=0$) and
$\Delta_0 \ll E_F$. A plot of the tunneling conductances
as a function of the bias voltage $eV$ for
different barrier strength $\chi$, shown in Fig.\ \ref{fig1},
confirms the $\pi$ periodic oscillatory behavior. The oscillation
amplitude is maximum for zero-bias ($eV=0$), as shown in Fig.\ \ref{fig2}, and
vanishes at the gap edge ($eV=\Delta_0$). From Fig.\ \ref{fig1}, we find two noteworthy
features. First, the tunneling conductance at the gap edge reaches a
value close to $2G_0$ independent of the barrier strength, and
second, the subgap tunneling conductance becomes close to $2G_0$
when $\chi=(n+1/2)\pi$ for any integer $n$ and any $eV \le \Delta_0$.

Both of the above-mentioned features can be understood by noting that when
$eV \ll E_F$, $\alpha \simeq  -\alpha'\simeq \gamma$
(Eq.\ \ref{wavenorm}). In this limit, using Eqs.\ \ref{coeff1} and
\ref{coeff2}, it can be shown that the reflection amplitude, for $eV \le \Delta_0$,
becomes $ r \simeq {\mathcal N}
\left[\cos\left(\chi\right)+ i \sin(\chi) \cos(\alpha)\right]/{\mathcal D}$,
where  ${\mathcal D} = \left[\cos(\alpha) \left[\cos(\beta)\cos(2\chi)+i \sin(\beta)\right]
+ 2 \sin(2\chi) \sin(\beta) \sin^2(\alpha)\right]$ and
\begin{eqnarray}
{\mathcal N} &=& 2 \sin(\alpha) \left[\sin\left(\chi+\beta\right)
-\sin\left(\chi-\beta\right)\right]. \label{rlimitn}
\end{eqnarray}
Note that ${\mathcal N}$ and hence $r$ vanishes and when $\alpha=0$ or
$\sin(\chi + \beta) = \sin(\chi -\beta)$. The former condition ($\alpha=0$) is a
manifestation of the well-known Klein paradox \cite{klein1} and occurs
since scattering of normal-incident Dirac electrons from a potential barrier can not
change their chirality \cite{geim1}. This effect, however, is not manifested
easily in the tunneling conductance since $G$ receives contribution from
electron approaching the barrier with all possible incidence angles
$\alpha \le \alpha_c$. The latter equality ($\sin(\chi + \beta) = \sin(\chi -\beta)$ )
represents condition for transmission resonance ($r=0$ and $|r_A|=1$) of a graphene NIS
interface and has the solutions $\beta=0$ and $\chi=(n+1/2)\pi$.

Such transmission resonances occur for all barrier strengths $\chi$ and angle of
incidence $\alpha$ when $eV=\Delta_0$ ({\it i.e.} $\beta=0$). Consequently,
$G(eV=\Delta_0) \simeq 2G_0$ independent of the barrier strength, as is also
well-known for conventional NIS junctions \cite{tinkham1}.
The novel aspect of a graphene NIS junction comes from the second class
of solution of the transmission resonance condition: $\chi = (n+1/2)\pi$. At
these special values of the barrier, $G \simeq 2G_0$ for any subgap voltage as
long as $eV \ll E_F$ \cite{comment1}. This feature, clearly seen in
Figs.\ \ref{fig1} and \ref{fig2}, is in sharp contrast to conventional NIS junction where
$G(eV < \Delta_0)$ always decreases with increasing barrier strength
\cite{tinkham1}. For $\chi \ne (n+1/2)\pi$ or $eV \ne \Delta_0$, $r \ne 0$ and
consequently $|r_A| < 1$ so that $G < 2G_0$. In particular, $r$ reaches a maxima leading to
minimum value of subgap tunneling conductance for $\chi=n \pi$. Thus, in contrast to
conventional NS junctions, $G(eV=0) < 2G_0$ for $\chi=0$ in graphene NS junctions
as noted earlier in Ref.\ \cite{beenakker1}.

\begin{figure}
\rotatebox{0}{
\includegraphics*[width=7.0cm]{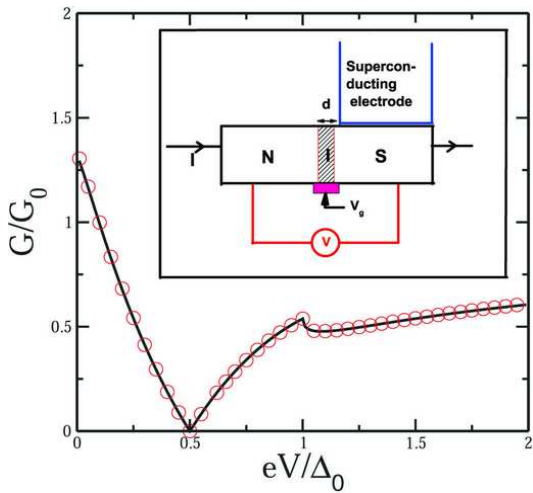}}
\caption{Tunneling conductance as a function of bias voltage
for effective barrier strengths
$\chi=0$ (solid line) and $\chi=\pi/2$ (open circle) for $U_0=25 E_F$
and $\Delta_0=2 E_F$. The tunneling conductance becomes barrier independent
and vanishes for $eV=E_F$. The inset shows a schematic experimental setup.
The dashed region sees a variable gate (shown as pink filled region) voltage $V_0$
which creates the barrier. Additional gate voltage $U_0$, which may be applied on
the superconducting side, and the current source is not shown to avoid clutter.}
\label{fig3}
\end{figure}

Next, we briefly explore the regime where $\Delta_0 \ge E_F$
and $U_0 \gg E_F$, so that $\Delta_0 \ll (U_0 + E_F)$
\cite{beenakker1}. Here the tunneling conductance, shown in Fig.\
\ref{fig3} for $\Delta_0=2E_F$ and $\chi=0,\pi/2$, becomes
independent of barrier strength. This can be intuitively understood
from the fact that a large Fermi surface mismatch acts as an
effective barrier for the electrons tunneling at the interface
\cite{tinkham1} which makes the presence of an additional
barrier irrelevant. Further, as seen from Fig.\ \ref{fig3}, the
tunneling conductance vanishes for $eV=E_F=0.5\Delta_0$ due to the
absence of Andreev reflection since $\alpha_c=0$ at this
bias-voltage. Our results in this limit therefore becomes identical
to those for NS junction studied in Ref.\ \cite{beenakker1}.

Finally, we discuss possible experimental setup (shown in inset of Fig.\ \ref{fig3})
to test our predictions. In the experiment, the local barrier can be
fabricated using methods of Ref.\ \cite{nov2}. The easiest
experimentally achievable regime corresponds to $\Delta_0 \ll E_F$
with aligned Fermi surfaces for the normal and superconducting
regions. We suggest measurement of tunneling conductance curves at
zero-bias ($eV=0$) in this regime. Our prediction is that the
zero-bias conductance will show an oscillatory behavior with change
of effective bias voltage with maxima(minima) when $\chi$ becomes
odd(even) integer multiples of $\pi/2$ . In graphene, typical Fermi
energy is $E_F \simeq 80$meV and the Fermi-wavelength is $\lambda =
2\pi/k_F \simeq 100$nm \cite{geim1,nov2}. For realization of the
thin and sharp barriers discussed in this work, one needs $d/\lambda
<< 1$ and $ V_0/E_F \gg 1$. Effective barrier strengths of
$500-1000$meV and barrier widths of $d \simeq 20-10$ nm, which can
be achieved in realistic experiments \cite{geim1,nov2}, shall
therefore meet the demands of the proposed experimental setup. To
observe the oscillatory behavior of the zero-bias tunneling
conductance, it would be necessary to change $\chi$ in small steps $\delta \chi$.
For barriers of a fixed width, with values of $d/\lambda=0.1$ and
$V_0/E_F=10$, this would require changing $V_0$ in steps of
approximately $12$meV which corresponds to $\delta \chi=0.1$.

In conclusion, we have a presented a theory of tunneling
conductance of graphene NIS junctions. We have demonstrated that
the tunneling conductance exhibits novel $\pi$ periodic oscillatory
behavior as a function of barrier strength of the junction and have
suggested experiments to observe this effect. The authors thank S.M.
Bhattacharjee, A. Ghosh, P.K. Mohanty,M. Khan, T. Senthil and V. B. Shenoy
for valuable discussions and Graduate Associateship Program at Saha Institute
which made this work possible.


\begin{thebibliography}{99}
\bibitem{comment1} Note that for $eV=0$, the equality $G=2G_0$
is exact since $\alpha=-\alpha'=\gamma$ at zero-bias.
\end{thebibliography}
\end{document}